# Swept Source Lidar: simultaneous FMCW ranging and nonmechanical beam steering with a wideband swept source

**MASAYUKI OKANO**[*] **AND CHANGHO CHONG**

*Santec Photonics Laboratories, Santec Corporation, 2350 Mission College Blvd, Santa Clara, CA 95054, USA*
*\*mokano@santec.com*

**Abstract:** Light detection and ranging (lidar) has long been used in various applications. Solid-state beam steering mechanisms are needed for robust lidar systems. Here we propose and demonstrate a lidar scheme "Swept Source Lidar" that allows us to perform frequency-modulated continuous-wave (FMCW) ranging and nonmechanical beam steering simultaneously. Wavelength dispersive elements provide angular beam steering while a laser frequency is continuously swept by a wideband swept source over its whole tuning bandwidth. Employing a tunable vertical-cavity surface-emitting laser and a 1-axis mechanical beam scanner, three-dimensional point cloud data has been obtained. Swept Source Lidar systems can be flexibly combined with various beam steering elements to realize full solid-state FMCW lidar systems.

## 1. Introduction

Light detection and ranging (lidar) is becoming ubiquitous for distance measurements in various applications, including autonomous driving, robotics, and three-dimensional (3D) sensing [1]. Using two-dimensional (2D) beam steering of an outgoing beam, objects can be detected in 3D space. Mechanical beam steering systems based on mature technologies, such as rotors, galvo scanners, polygon mirrors, and micro-electro-mechanically-systems (MEMS) mirrors, have been utilized in a wide range of lidar applications. In recent years, solid-state lidar systems, which have no moving components, have attracted more attention in industries where robustness and stability are critical. Various nonmechanical beam steering methods have been studied; however, these methods still have limited performances or rely on complex immature technologies, including silicon-photonics optical phased arrays (OPAs) [2], liquid-crystal waveguides [3], vertical-cavity surface-emitting laser (VCSEL) based waveguides [4], and photonic-crystal waveguides [5]. Thus, simple solid-state beam steering mechanisms are desired for the realization of robust and stable lidar systems.

In terms of light detection methods, time-of-flight (TOF) ranging based on detections of pulsed light has been the dominant technology especially in automotive lidar applications [6]. Alternatively, there has been a growing interest in frequency-modulated continuous-wave (FMCW) ranging based on optical heterodyne detection because FMCW ranging has technical advantages over TOF ranging [7] as follows: First, it can have high sensitivity, large dynamic range, and high range resolution [8]. Therefore, targets of varying reflectivity can be detected over long distances within eye-safe regulations [9]. Second, it is inherently immune to ambient light, such as the sunlight and light from other lidar systems. In contrast, ambient light can reduce signal-to-noise ratio (SNR) and cause false detections in TOF ranging [10,11]. Finally, the instantaneous velocity of targets can be measured.

A solid-state FMCW lidar system presents an attractive solution for a wide range of applications. However, there are challenges to overcome as follows: First, it is difficult to combine two state-of-the-art techniques, a sophisticated FMCW ranging and a solid-state beam steering mechanism, into one simple integrated solution. Second, a high-performance swept source with a very narrow linewidth is required for FMCW ranging, where the detection range is limited by its coherence length.

Here we propose [12] and demonstrate a solid-state FMCW lidar scheme, which we call "Swept Source Lidar [13]". Swept Source Lidar can perform FMCW ranging and solid-state beam steering simultaneously. Wavelength dispersive elements provide nonmechanical beam steering while the laser frequency of a wideband swept source is continuously swept over its whole tuning bandwidth. Therefore, Swept Source Lidar systems can eliminate the need to steer beams along at least one of two axes for 2D beam steering. Our lidar systems can be simpler than other solid-state FMCW lidar systems because FMCW ranging and solid-state beam steering are unified into a single process performed during a single monotonic frequency sweep. Owing to its simplicity, Swept Source Lidar systems can be combined with various 1-axis beam steering mechanisms. Since the beam steering range is intrinsically proportional to the tuning bandwidth, a wideband swept source is suitable.

To demonstrate our scheme, we have adapted our swept-source optical coherence tomography (SS-OCT) technology developed for medical and industrial applications [14]. Since SS-OCT and FMCW ranging are both based on optical coherent detection [1,15], common techniques can be applied [16]. As a wideband swept source, we employ our tunable VCSEL [17]. Tunable VCSELs have recently been used in long-range distance measurements [18,19] and the coherence lengths can be over several hundred of meters [17,20]. In SS-OCT and FMCW lidar systems, the instantaneous linewidth during frequency sweep can be obtained by coherence length measurement rather than direct measurement such as self-heterodyne method. The coherence length is measured by the optical path length difference of a Mach-Zehender interferometer at which the interference fringe amplitude drops by 3 dB [17,20]. Our electrically-pumped tunable MEMS-VCSEL (HSL-1, Santec) has a long coherence length (> 150 m) corresponding to its narrow instantaneous linewidth (< 1 MHz), a wide tuning bandwidth (> 70 nm), and a variable sweep rate (up to 400 kHz). More recently, Y. Zhai *et al.* have demonstrated a concept similar to our scheme [21]. However, their experiment was performed at a slow sweep rate and limited to 2D measurements. In the present work, we demonstrate our scheme using our tunable MEMS-VCSEL at a sweep rate of as fast as 10 kHz with a tuning bandwidth of 40 nm at a center wavelength of 1060 nm. The laser beam from the tunable VCSEL is 2D raster-scanned using nonmechanical beam steering along the fast axis over 8 deg and mechanical beam steering with a galvo scanner along the slow axis. As a result, 3D point cloud data for a target placed at 0.5 m has been successfully obtained. In order to detect targets for longer range, we run the tunable VCSEL at a slower sweep rate of 300 Hz and 3D point cloud data for targets placed at 5 m has been obtained.

## 2. Theory

### 2.1 Conventional FMCW lidar

In this section, a conventional FMCW lidar system is introduced. A general architecture of the system is depicted in Fig. 1. The system consists of a swept source and a beam steering device. The outgoing beam from the swept source is steered by the beam steering device as shown in Fig. 1(a). At each beam emission angle $\theta_i$ ($i = 1,…,M$) of $M$ scanning points, a laser frequency $f(t)$ is swept in time $t$, and a fraction of the outgoing beam can return by bouncing at targets with a round-trip time $\Delta t_i$. Then, the distance to targets $R_i$ at each beam angle is obtained by the beat frequency $\Delta f_i$ of the interference between the outgoing beam and the reflected beam. When the laser frequency $f(t)$ is linearly swept as expressed in the following equation:

$$f(t) = \frac{F}{T}t + f_0, \tag{1}$$

where $f_0$ is the initial frequency, $F$ is the tuning bandwidth, and $T$ is the tuning period, the range $R_i$ at each beam angle is obtained by

$$R_i = \frac{cT}{2F} \Delta f_i, \quad (2)$$

where $c$ is the speed of light, taking into account the round-trip time $\Delta t_i$ [8]. The range $R_i$ can also be identified using the Fast Fourier Transform (FFT) of the interference signal. The position of the peak standing above the noise floor in the FFT signal corresponds to the distance to the targets, as illustrated in Fig. 1(b). In the SS-OCT framework, a single frequency sweep for ranging and the FFT signal obtained from an interference signal are called "A-scan" and "point spread function (PSF)", respectively [14]. In this way, a 3D point cloud as 3D lidar data for targets is obtained using 2D beam steering over all scanning points. In conventional SS-OCT systems, 3D SS-OCT data is obtained as a stack of PSFs over 2D scanning points using 2D transversal beam scans called "B-scan" and "C-scan" [14]. Therefore, 2D beam steering is required to obtain 3D point cloud data for conventional FMCW lidar systems.

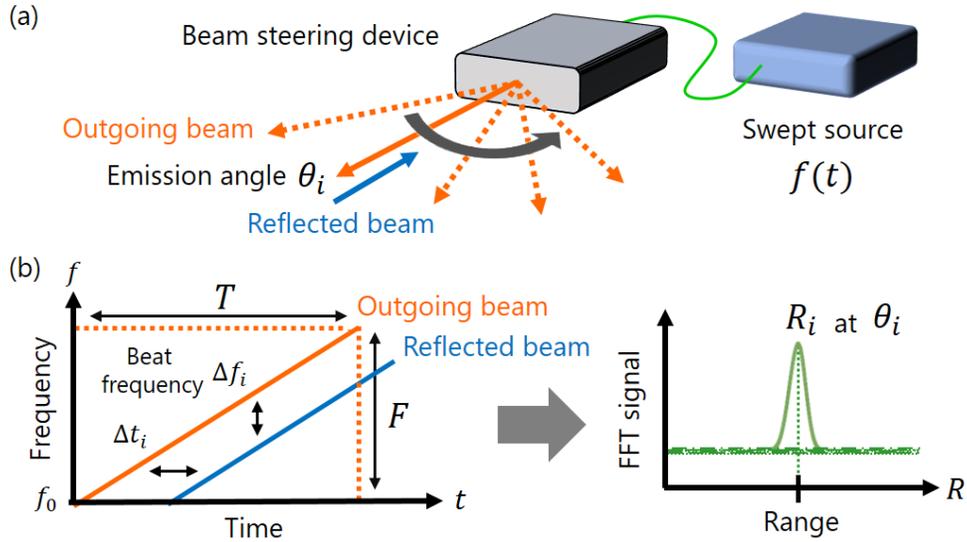

Fig. 1. (a) Schematic of a conventional FMCW lidar system, consisting of a swept source and a beam steering device. (b) The range for targets at each beam emission angle can be determined from the beat frequency or the FFT signal obtained from the interference between the outgoing and reflected beams.

Given the linear ramp of the laser frequency described in Eq. (1), the range resolution $\Delta R$ is given by [22]

$$\Delta R = \frac{c}{2F}. \quad (3)$$

Therefore, a larger tuning bandwidth provides a higher range resolution. Note here that the depth resolution for SS-OCT $\Delta R_{oct}$ is generally written in the following equation [22,23]:

$$\Delta R_{oct} = \frac{2\ln 2}{\pi} \frac{\lambda_c^2}{\Delta \lambda} \simeq 0.44 \frac{c}{\Delta \nu}, \quad (4)$$

where $\lambda_c$ is the center wavelength, $\Delta \lambda$ and $\Delta \nu$ are the full width at half maximum (FWHM) of the light source in the wavelength and frequency domains, respectively. Though these two range resolutions have slightly different values depending on the frequency sweep assumption, they are in the same order. When the range $R_i$ is obtained from FFT signals, the range resolution can be estimated by the FWHM of its peak. The range resolution can be degraded from the theoretical resolution because of the size of FFT bins. The maximum detection range $R_{max}$ is limited by the sampling rate $B$ of a data acquisition (DAQ) system when the detection system has sufficiently large electronic bandwidth. Because the detection bandwidth is limited to half the sampling rate considering the Nyquist theorem [24], $R_{max}$ is given by [22]:

$$R_{max} = \frac{BcT}{4F}. \qquad (5)$$

As Eq. (5) indicates, a higher sampling rate $B$ or a slower frequency ramp rate $F/T$ provides a longer maximum detection range. Regarding emission angles, the angular resolution can be determined by 2D beam steering processes, depending on the step size of the scan and beam properties such as beam size, beam pattern, and beam divergence.

## 2.2 Swept Source Lidar

This section introduces our scheme called "Swept Source Lidar". Its basic setup is illustrated in Fig. 2. A Swept Source Lidar system consists of a swept source and a wavelength dispersive element, instead of a beam steering device used in conventional FMCW lidar systems. Because of the dispersive effect, a laser beam is nonmechanically steered while the laser frequency is swept. The emission angle of the outgoing beam $\theta^{ss}(t)$ changes over time depending on the frequency sweep $f(t)$. Throughout the paper, we refer to this beam steering process as "swept source scan". During a single monotonic frequency sweep, an interference signal is detected continuously. Swept Source Lidar then divides the obtained interference signal $I_s(t)$ into multiple segments:

$$\bigcup_{i=1}^{N}[t_{i-1}, t_i], \qquad (6)$$

where $N$ is the number of segments, $t_{i-1}$ and $t_i$ are the initial and final times of $i$th segment, respectively. At each beam angle $\theta_i^{ss}$, the range for targets $R_i^{ss}$ is obtained directly by

$$R_i^{ss} = \frac{cT}{2F}\Delta f_i^{ss}, \qquad (7)$$

where $\Delta f_i^{ss}$ is the beat frequency of the interference between the outgoing and reflected beams.

The range $R_i^{ss}$ can also be identified using the FFT of the interference signal at each segment $I_s([t_{i-1}, t_i])$. The peak position in the FFT signal corresponds to the distance to objects in the physical space. When the light return from multiple targets within a segment, multiple peaks appear in the FFT signal at positions corresponding to the distance to targets. The highest peak among them can be selected as a lidar data point. Since the dispersive element is static, delayed light from another segment at a beam angle $\theta_j$ (j < i) can return and couple through the dispersive element to the interference signal of the $i$th segment. The peaks of this light can be distinguished in the FFT signal because their beat frequencies are apparently outside of the expected range. They can be eliminated using low-pass filtering at the data processing stage. Alternatively, these intersegment coupling from different angles can be avoided by assigning the swept source scan to the slow axis of 2D beam steering, as will be discussed in Sec. 4.4. Ranges over $N$ segments are obtained while the laser frequency is continuously swept over the whole tuning bandwidth. Thus, FMCW ranging and nonmechanical beam steering are unified into a single process that can provide a 2D point cloud as 2D lidar data during a single frequency sweep. In SS-OCT terminology, an interference signal obtained by a single A-scan is divided into segments and a 2D SS-OCT image is obtained without additional 1-axis beam steering for B-scan, which is required in conventional SS-OCT systems [14]. By using swept source scans along two axes or combining a 1-axis swept source scan with additional 1-axis beam steering, a 3D point cloud as 3D lidar data over 2D beam steering is obtained.

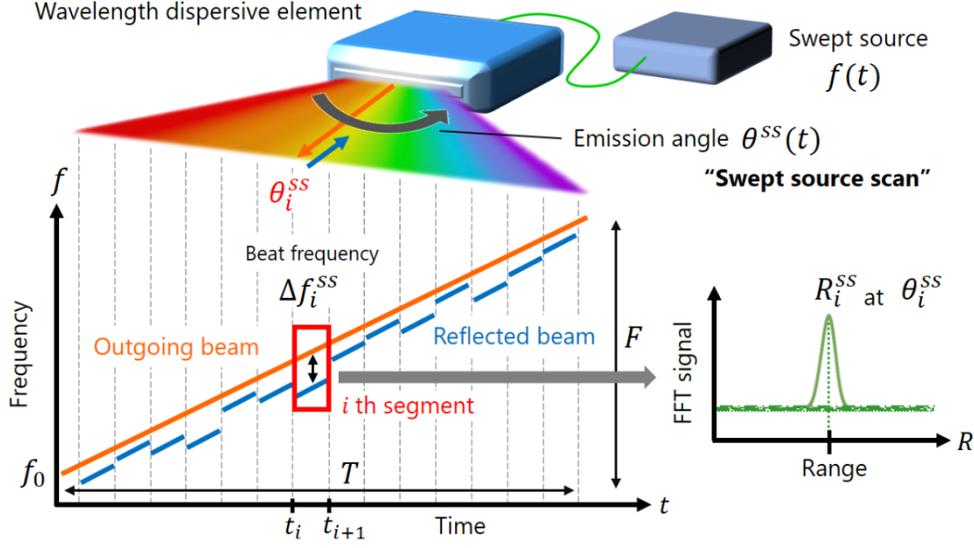

Fig. 2. Schematic of "Swept Source Lidar" system, consisting of a swept source and a wavelength dispersive element. Because of the dispersive effect, the laser beam from the source is nonmechanically steered while the laser frequency is continuously swept as "swept source scan". The range for targets at each beam emission angle can be determined from the beat frequency or the peak position in the FFT signal at each segment.

The range resolution $\Delta R_i^{ss}$ at each segment is expressed in the form:

$$\Delta R_i^{ss} = \frac{c}{2F_i}, \tag{8}$$

where $F_i = f(t_i) - f(t_{i-1}) = (F/T) \times (t_i - t_{i-1})$ is the tuning bandwidth for the *i*th segment. When an interference signal is divided into $N$ segments uniformly, the range resolution for each segment is given by $N\Delta R$, which is a factor of $N$ lower than $\Delta R$. The angular pixel size $\Delta \theta_i^{ss}$ for the *i*th segment is $\Theta^{ss}/N$, where $\Theta^{ss} = |\theta^{ss}(t_N) - \theta^{ss}(t_0)|$ is the whole angular beam steering range. These relations show that there is a tradeoff between the range resolution $\Delta R_i^{ss}$ and the angular pixel size $\Delta \theta_i^{ss}$. In lidar applications, the range resolution does not necessarily define the measurement accuracy, which can be several orders of magnitude better than the range resolution, unlike in typical OCT imaging applications where the depth resolution directly affects the image quality [1,15]. The number of segments can be set large enough to achieve a fine angular pixel size by reducing a range resolution. The range resolution can be degraded from the theoretical resolution because of the range pixel size $\Delta Z_{FFT}^{ss}$, defined as the FFT bin size, and the angular resolution can be degraded from the angular pixel size $\Delta \theta_i^{ss}$, depending on beam properties such as the beam size. It is important to note that the maximum detection range $R_{max}^{ss}$ is the same as conventional FMCW lidar systems $R_{max}$ because it is determined by the frequency ramp rate $F/T$ and independent of $N$.

## 3. Experimental results

### 3.1 Experimental setup

A schematic diagram of the experimental setup is illustrated in Fig. 3. A tunable MEMS-VCSEL (HSL-1, Santec) was employed as a swept source [17]. The bi-directional frequency sweep rate was set to 10 kHz and the wavelength tuning bandwidth was about 40 nm centered at 1060 nm. Its corresponding frequency tuning bandwidth $F$ is 11 THz. The output light from the swept source is introduced into two fiber-based optical interferometers. The first one is a "k-clock" interferometer used to linearize frequency sweeps in data processing. Such a k-sampling (resampling) technique has been used in our SS-OCT framework [14,25]. The optical path delay in this Mach-Zehnder interferometer was 43 mm in optical fiber length. Interference signals are detected with an InGaAs balanced photo detector (BPD) with a bandwidth of 1 GHz (PDB481-AC, Thorlabs). The other interferometer is the main interferometer used to measure distances to targets. The outgoing beam from an optical fiber was loosely focused at targets with an adjustable fiber collimator (ZC618APC-C, Thorlabs). Its averaged optical power was 1 mW and the beam diameter was 3 mm at the collimator. A diffraction grating (33009FL01-530R, Newport) with a groove density of 1200 grooves/mm was placed at 50 cm from the collimator just after a 1-axis galvo scanner. The galvo scanner projects and deflects the beam in the vertical axis towards the grating as the "galvo scan". The first-order diffracted beam from the grating is steered horizontally while the laser frequency is swept. The polarization of the beam to the grating was controlled with a polarization controller to be vertical (S-polarization) to maximize the first-order diffraction efficiency up to 89%. The zeroth-order diffraction with an efficiency of less than 11% was blocked. Thus, the outgoing beam was steered by the swept source scan and the galvo scan in the fast and slow axes of 2D raster scanning, respectively.

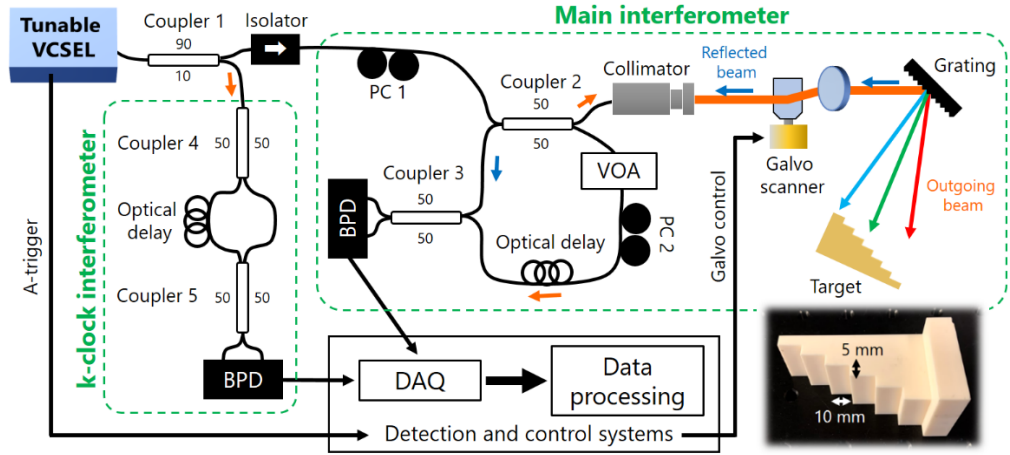

Fig. 3. Experimental setup for Swept Source Lidar demonstration. VCSEL: Vertical-cavity surface-emitting laser, PC: Polarization controller, VOA: Variable optical attenuator, BPD: Balanced photodetector, and DAQ: Data acquisition system. Picture of the target is shown on the bottom right corner of the figure.

A ceramic target with step structure was placed at 0.5 m from the grating. A fraction of the beam reflected by the target can be coaxially recoupled into the interferometer and then interfere with the reference beam. This coaxial configuration has been commonly used in SS-OCT systems [14,15]. The reference optical power was controlled with a variable optical attenuator to optimize the SNR of signals. The optical delay was introduced into the reference path, and optical path lengths for two arms of the main interferometer were balanced at 80 cm from the collimator, about 20 cm before the ceramic target. Interference signals are detected with an InGaAs BPD (WL-BPD1GA, Wieserlabs) with a bandwidth of 1 GHz. All data are obtained by using a DAQ board (APX-5200B, Aval Data) with a resolution of 12 bits and a sampling rate of 1 GS/s. Sampling the data for each frequency sweep is triggered by a TTL A-trigger signal from the swept source. Signal detection and signal processing are performed with a field-programmable gate array (FPGA) on the DAQ board and a LabVIEW program.

When a beam is incident on a grating, the diffraction grating relation is given by [26]

$$m\lambda = d(\sin\theta_{in} + \sin\theta_{out}), \tag{9}$$

where $m$ is the diffraction order, $d$ is the grating period, $\theta_{in}$ and $\theta_{out}$ are the input and output beam angles, respectively. Substituting Eq. (1) into Eq. (9) gives the emission angle of an outgoing beam $\theta^{ss}(t)$ in the swept source scan:

$$\theta^{ss}(t) = \sin^{-1}\left(\frac{mc}{f(t)d} - \sin\theta_{in}\right). \tag{10}$$

The emission angle can then be approximated using Taylor expansion by

$$\theta^{ss}(t) \approx \theta^{ss\prime}(0)t + \theta^{ss}(0) = -\frac{mc}{f_0^2 d\sqrt{1-\left(\frac{mc}{f_0 d} - \sin\theta_{in}\right)^2}}\frac{F}{T}t + \theta_0^{ss}, \tag{11}$$

where $\theta_0^{ss} = \theta^{ss}(0)$ is the initial emission angle. Under this approximation, the whole angular beam steering range is proportional to $F$. Therefore, a larger tuning bandwidth provides a wider angular beam steering range. In the present work, the diffraction order $m = 1$, the grating period $d = 1/1200$ mm, the incident angle $\theta_{in} = 20$ deg, and the tuning period $T$ for the upward frequency ramp in bi-directional sweeps was 35 μs. The calculated beam steering range and the beam steering rate $|d\theta^{ss}/dt|$ are 8 deg and $1.9 \times 10^5$ deg/s, respectively.

### 3.2 Data processing

A schematic illustration of data processing with sample data is shown in Fig. 4. First, an interference signal obtained from the main interferometer $I_s(t)$ is rescaled to the frequency domain $S(f)$ with the constant frequency intervals using the corresponding interference signal (k-clock signal) obtained from the k-clock interferometer $I_k(t)$ [14,25]. The observed bias in the k-clock signal is caused by the asymmetry and wavelength dependence of the k-clock interferometer and the balanced detection. The k-clock signal frequency typically varies during each sweep because of the inherent nonlinearity in frequency sweeps (Fig. 4). Then, the resampled interference signal $S(f)$ is divided into $N$ segments as defined by the segmentation:

$$\bigcup_{i=1}^{N}[f_{i-1}, f_i], \tag{12}$$

where $f_{i-1}$ and $f_i$ are the initial and final frequencies for the $i$th segment, respectively. For simplicity, the interference signal was divided uniformly into 30 segments ($N = 30$). The tuning bandwidth for each segment was set to the same value $F_i$ of $F/N = 360$ GHz, where $F = f_N - f_0$ is the whole tuning bandwidth of 11 THz. The range resolution $\Delta R_i^{ss}$ calculated from the tuning bandwidth for each segment is 0.42 mm. Both $I_s(t)$ and $I_k(t)$ were obtained at the sampling rate of 1 GS/s and the resampled interference signal $S(f)$ had a total resampling points of 61,440. The resampled signal for each segment $S([f_{i-1}, f_i])$ had 2,048 ($= 2^{11}$) resampling points $N_R$ and the corresponding FFT signal had 1,024 ($= 2^{10}$) pixels. Then, the range for targets at each segment is determined by measuring the peak position in the FFT signal. The whole detection range for each FFT signal $Z_{FFT}^{ss}$ is given by [14,22]

$$Z_{FFT}^{ss} = \frac{cN_R}{4F_i} = \frac{cN_R N}{4F}. \tag{13}$$

The calculated whole detection range is 0.43 m, which is in good agreement with the measured value of 0.45 m. The range pixel size $\Delta Z_{FFT}^{ss}$ calculated by the measured whole detection range is 0.44 mm/pixel. The range resolution estimated by the FWHM of the peak in each FFT signal was 1.3 pixels = 0.56 mm when the k-clock signal delay to the interference signal was optimized. The measured resolution is in close agreement to the theoretical resolution $\Delta R_i^{ss}$ as the resampling compensates the nonlinearity in frequency sweeps. The maximum detection range $R_{max}^{ss}$ calculated from the sampling rate $B$ of 1 GS/s is 0.25 m.

A 2D density plot is obtained from all of the FFT signals for 30 segments. Each FFT signal is colored according to intensity in the plot (Fig. 4). In SS-OCT terminology, a 2D SS-OCT image is obtained from a single A-scan without a B-scan. The calculated angular pixel size at each segment $\Delta \theta_i^{ss}$ is 0.27 deg/pixel. The angular resolution can be degraded from the angular pixel size because of the beam size. The FWHM of the intensity profile of the Gaussian beam focused at the target was measured to be 0.19 mm. Its corresponding angular width is 0.022 deg, which is one order of magnitude smaller than the angular pixel size. Next, a 2D peak plot as a 2D point cloud is obtained by detecting the peak position in each FFT signal. The 2D step structure of the target has been successfully captured (Fig. 4). Full data processing can be fast like SS-OCT because the advantage of FFT processing remains.

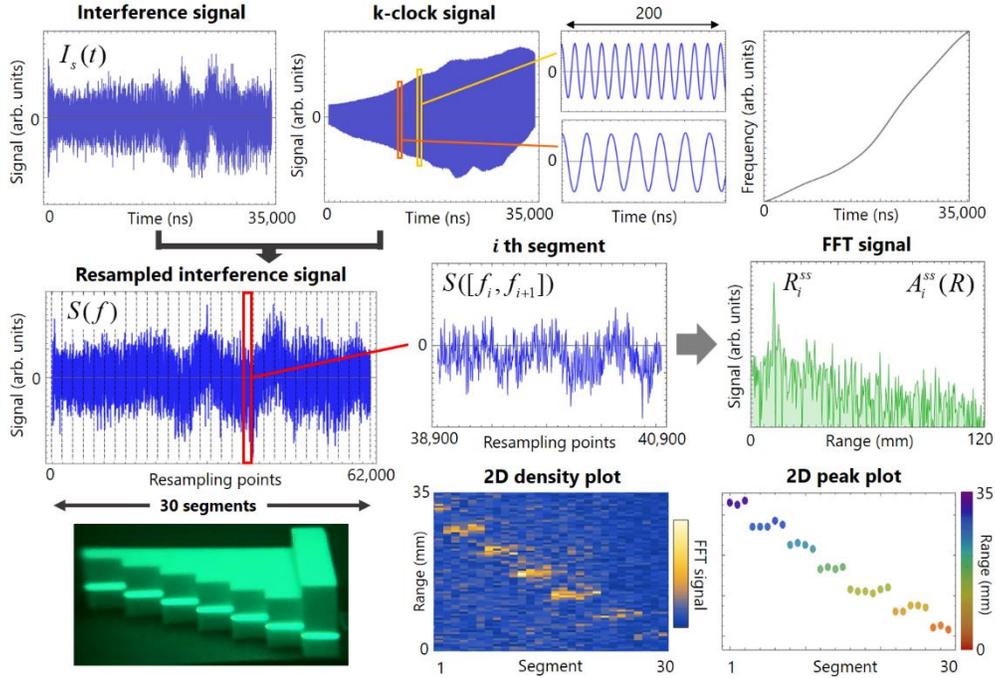

Fig. 4. Data processing in Swept Source Lidar system. The interference signal taken during a single frequency sweep is resampled with a constant interval in the frequency domain using the corresponding k-clock signal. The frequency sweep curve is shown on the top right corner of the figure. Next, the resampled signal is divided into segments. The range at each segment can be obtained from the FFT of the resampled interference signal. A 2D density plot is obtained from FFT signals over all segments. A 2D peak plot is provided by peak detection. Picture of the object with a beam steered in the horizontal axis is on the bottom left corner of the figure.

*3.3 Demonstration of 3D Swept Source Lidar*

Swept Source Lidar systems can be combined with various beam steering mechanisms. As an additional beam steering mechanism, a galvo scanner, which has been widely employed in SS-OCT systems [14,15], was used to perform Swept Source Lidar in 3D. The swept source scan and the galvo scan were used for the fast and slow axes of 2D raster scanning, respectively, as is illustrated in Fig. 5(a). As a picture taken with an infrared camera is shown in Fig. 5(b), the beam was 2D raster-scanned on the target. The galvo scan had 100 scan lines in total over 2.9 deg along the vertical axis and the corresponding angular pixel size is 0.029 deg/pixel. The galvo scanner was driven by triangular waves and triggered synchronously with A-trigger signals by the control system.

Figure 5(c) shows a depth map of the obtained 3D point cloud. The beam was steered along the horizontal axis over 8.0 deg for 30 segments by the swept source scan and the vertical axis by the galvo scan. The 3D point cloud was colored according to range. The 3D structure of the target has been successfully captured. The 2D step structure has been detected at each galvo scan. The frame rate can be calculated from the frequency sweep rate divided by the number of galvo scan lines. The frame rate was approximately 100 Hz with negligible data processing times. The 3D Swept Source Lidar has been successfully demonstrated at a high frame rate with additional 1-axis mechanical beam steering. It is important to point out that only 1-axis additional beam steering was used to perform 3D FMCW ranging, although 2D beam steering is required in conventional 3D FMCW lidar systems.

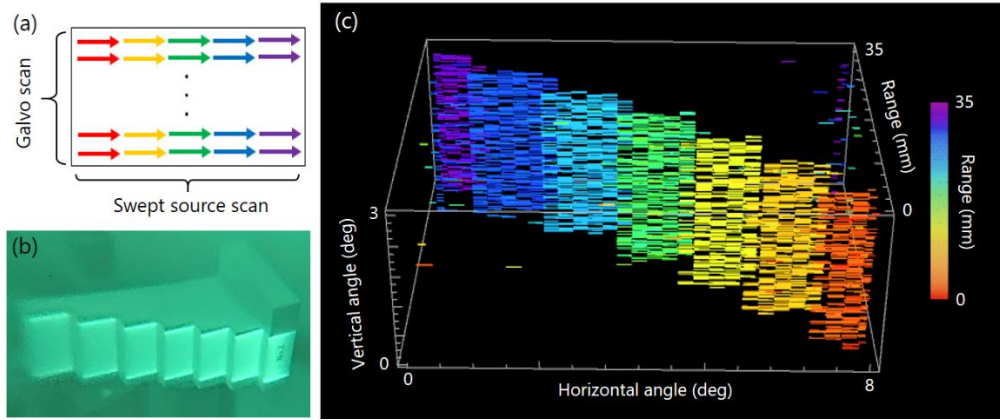

Fig. 5. (a) Schematic of 2D raster scanning with the swept source scan along the horizontal axis and the galvo scan along the vertical axis. (b) Picture of a target placed at 0.5 m from the grating with a laser beam 2D raster-scanned. (c) Depth map of the 3D point cloud for the target.

### 3.4 Long-range 3D Swept Source Lidar

In order to detect targets over longer distances, the maximum detection range $R_{max}^{ss}$ needs to be longer. The maximum detection range was increased to 12 m by slowing our tunable MEMS-VCSEL down to 300 Hz with a tuning period $T$ of 800 μs and a tuning bandwidth $F$ of 5.0 THz. The optical delay in the k-clock interferometer was set to 2 m in optical fiber length to maximize the SNR of signals. Two sheets of white paper were placed 300 mm apart as targets at 5 m from the grating. Characters of "santec" were cut out of the sheet placed nearside as in the picture shown in Fig. 6(a). The beam was 2D raster-scanned by the swept source scan in the vertical fast axis and the galvo scan in the horizontal slow axis. The swept source scan over 1.9 deg with 45 segments and the galvo scan over 5.0 deg with 200 scan lines were performed. The calculated vertical and horizontal angular pixel sizes are 0.042 and 0.025 deg/pixel, respectively. The angular resolution can be degraded because of the beam size. The FWHM of the intensity profile of the focused Gaussian beam was measured to be 1.2 mm. Its corresponding angular width is 0.012 deg, which is less than a half of the angular pixel size. The frame rate in the measurements was approximately 1.5 Hz.

The tuning bandwidth of each segment $F_i$ was 110 GHz and the calculated range resolution $\Delta R_i^{ss}$ is 1.4 mm. The resampled interference signal for each segment had 8,192 (= $2^{13}$) resampling points $N_R$ and its corresponding FFT signal had 4,096 pixels. The calculated whole detection range $Z_{FFT}^{ss}$ is 5.6 m, which is in agreement with the measured value of 5.6 m. The range pixel size $\Delta Z_{FFT}^{ss}$ calculated by this value is 1.4 mm/pixel. The range resolution estimated by the measured FWHM of a peak in each FFT signal was 1.2 pixels = 1.6 mm when the k-clock signal delay was optimized. This is close to the theoretical value and is comparable to the calculated range pixel size. The 3D point cloud data has been obtained and is shown as a depth map with pixels colored according to range (Fig. 6(b)) and a height map with dots colored according to vertical angle (Fig. 6(c)). The characters "santec" have been successfully captured. The long-range 3D Swept Source Lidar for the targets placed at 5 m has been successfully demonstrated with only 1-axis beam steering.

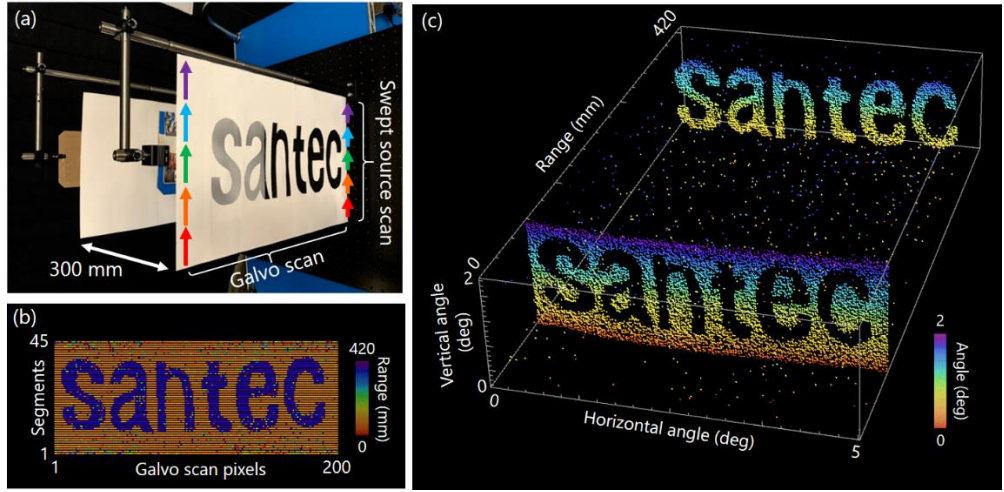

Fig. 6. (a) Picture of targets (two sheets of white paper) placed at 5m from the grating. The beam was 2D raster-scanned by the swept source scan along the vertical fast axis and the galvo scan along the horizontal slow axis. (b) Depth map of a 3D point cloud with 45 segments in the swept source scan and 200 pixels in the galvo scan. (c) Height map of the 3D point cloud of the targets.

## 4. Discussions

### 4.1 Comparison of FMCW lidar schemes

This section compares Swept Source Lidar to various FMCW lidar schemes as depicted in Fig. 7. In a conventional FMCW lidar system, a beam from is steered by a beam steering device, and FMCW ranging is performed at each beam scanning point (Fig. 7(a)). A 2D beam steering mechanism and a complex FMCW ranging system are both required to obtain a 3D point cloud. Recently, 2D beam steering using a one-dimensional OPA and grating waveguides as dispersive elements has been studied [27,28] and demonstrated FMCW ranging [29,30]. An outgoing beam is steered nonmechanically using the dispersive effect (Fig. 7(b)). In this scheme, the beam is steered discretely and FMCW ranging is performed by modulating the laser frequency locally at each beam steering angle, requiring a complex frequency sweep system.

Compared with these systems, Swept Source Lidar performs both FMCW ranging and nonmechanical beam steering simultaneously during a single monotonic frequency sweep without additional laser frequency modulators (Fig. 7(c)). This technique of converting the frequency sweep to both beam steering and frequency modulation for heterodyne detection offers a simple solution of lidar configuration whilst maintaining the advantages of FMCW ranging. Owing to its simplicity, Swept Source Lidar systems can be combined with various 1-axis mechanical or solid-state beam steering mechanisms. Furthermore, there exist a variety of choices for wavelength dispersive elements, including diffraction gratings, prisms, liquid crystals, and virtually imaged phased arrays (VIPAs) [31]. To realize a full 3D solid-state FMCW lidar system, swept source scans can be used along both axes of 2D beam steering. For instance, two dispersive elements with orthogonal optic axes can be employed [32].

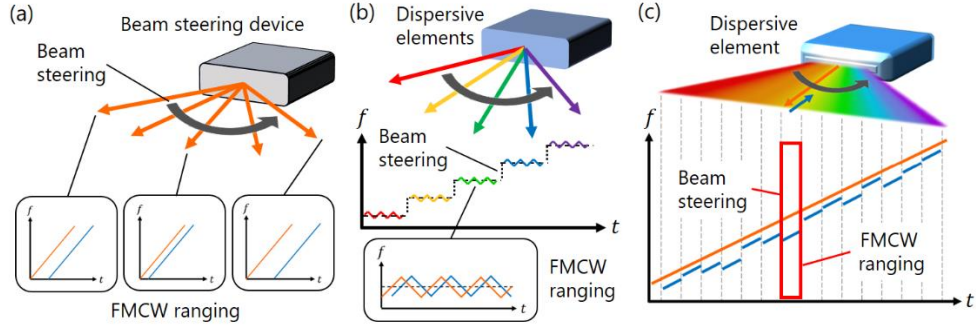

Fig. 7. Schematic overview of FMCW lidar systems. (a) Conventional FMCW lidar system with a beam steering device. (b) Beam steering system using dispersive elements and performing FMCW ranging at each beam steering angle discretely. (c) Swept Source Lidar system performing FMCW ranging and nonmechanical beam steering continuously and simultaneously using a wavelength dispersive element.

*4.2 SNR of FFT signals*

In lidar applications such as sensors for automotive applications, a long detection range of a few hundred meters is expected [1]. Generally, the detection range of FMCW ranging is limited by the coherence length and the SNR, in addition to the frequency ramp rate and the sampling rate. Since our tunable MEMS-VCSEL has a long coherence length in the order of the expected detection range or greater [17], the coherence length is not a limiting factor. In this section, the SNR in Swept Source Lidar systems is discussed. A series of SNR measurements are shown in Fig. 8. The SNR can be estimated by the peak height from the noise floor in each FFT signal with 50-time averaging. We swept the frequency of the tunable VCSEL at 300 Hz with the same parameters used in Sec. 3.4. A sheet of white paper was used as a target, the laser beam was loosely focused at it, and the FFT signals for the 1st segment of $N$ segments were used.

First, FFT signals with the target placed at several distances $L$ from the collimator were measured. The number of segments $N$ was 12, the number of resampling points $N_R$ at each segment was 32,768 ($= 2^{15}$), and the distance $L$ varied from 1.5 to 4.5 m. The 50-time averaged FFT signals in decibels are normalized to the highest value in all signals (colored lines, Fig. 8(a)). The k-clock signal delay was optimized for $L = 4.5$ m. The FFT signals show peaks at various positions corresponding to the physical positions of the target. The black line shows the 50-time averaged FFT signal without the target. The non-flat noise floor over the whole detection range is likely caused by the light scattering in the system and wavelength dependence of the interferometer. As the SNR with each target position is plotted (red dots, Fig. 8(b)), the measured SNR decreased as the distance $L$ increased. The SNR in SS-OCT [22,23] and FMCW lidar [33] systems is proportional to the optical power of reflected beams at the receiving aperture $P_r(L)$, which is given by

$$P_r(L) = \frac{\eta_t A_r \rho_T}{\pi L^2} P_0, \qquad (14)$$

where $P_0$ is the optical power at the transmitting aperture, $\rho_T$ is the reflectivity of a target, $A_r$ is the receiving aperture area, and $\eta_t$ is the transmittance of beams when the beam spot size is smaller than the target [33,34]. In Swept Source Lidar systems, $\eta_t$ can be estimated by the diffraction efficiency $\eta_g$ as $\eta_t = (\eta_g)^2$, assuming negligible loss in the air. As Eq. (14) indicates, the SNR decreases as $L$ increases. A theoretical fit using a power function (blue line, Fig. 8(b)) shows the SNR proportional to $L^{-1.9}$, which is in good agreement with the predicted dependence of $L^{-2}$. The SNR is also affected by beam parameters and k-clock signal delays.

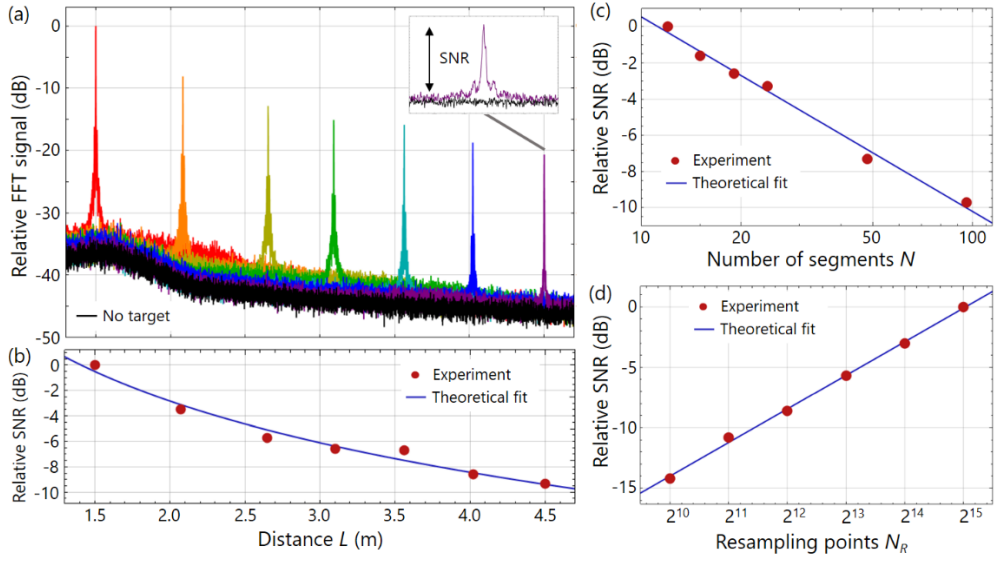

Fig. 8. (a) FFT signals with the target placed at several distances $L$ from the collimator (colored lines) and without the target (black line). The inset on the top right corner shows the FFT signals for $L = 4.5$ m. (b) Relative SNRs of the FFT signals with the distances $L$ (red dots) and a power function fit curve (blue line). (c) Relative SNRs with various $N$ (red dots) and a linear fit curve (blue line). (d) Relative SNRs with various $N_R$ (red dots) and a linear fit curve (blue line).

Second, the SNRs of FFT signals with various $N$ were measured. The target was placed at $L = 1.5$ m and $N_R$ for each segment was 32,768 (= $2^{15}$). The relative SNRs normalized to the highest value (red dots, Fig. 8(c)) decreased as $N$ increased. The SNR in Swept Source Lidar systems $SNR_{SSL}$ can be given as a function of $N$ by

$$SNR_{SSL} \propto \frac{P_r(L)T_d}{E_v} = \frac{P_r(L)T}{E_v N}, \quad (15)$$

where $E_v = h\nu$ is the single-photon energy at the laser frequency $\nu$ with Planck's constant $h$ and $T_d$ is the integration time [22,23,35]. Since the tuning period $T$ is divided into $N$ segments, $T_d = T/N$ and the SNR decreases as $N$ increases. A linear fit to the experimental data on the double logarithmic chart (blue line, Fig. 8(c)) shows the SNR is proportional to $N^{-0.97}$, which is in good agreement with the theoretical prediction of $N^{-1}$. Since the angular pixel size is determined by $\Theta^{ss}/N$, there is a tradeoff between the angular resolution and the SNR.

Third, the relative SNRs measured with various $N_R$ for each segment are plotted (red dots, Fig. 8(d)). The target position $L = 1.5$ m, $N = 12$, and $N_R$ varied from $2^{10}$ to $2^{15}$, which is limited by the FPGA code for signal processing. The result shows the SNR decreased as $N_R$ decreased. The number of sampling data points $N_S$ is $BT/N = 67,000$ on average and the maximum $N_R$ of 32,768 is about a half of the average $N_S$. As $N_R$ decreases, the SNR decreases at the same rate as a result of the reduced process gain of FFT processing [24]. A linear fit to the data on the double logarithmic chart (blue line, Fig. 8(d)) shows 2.8 dB/octave decrease in the SNR with decreasing $N_R$, which is consistent with the theoretical value of 3 dB. The SNR in SS-OCT systems can be written as a function of $N_S$ as

$$SNR_{SS} = \frac{N_S}{2} SNR_{TD}, \quad (16)$$

where $SNR_{SS}$ and $SNR_{TD}$ are SNRs in SS-OCT and Time-Domain OCT systems, respectively [22,23,35]. This is one of the advantages of SS-OCT. However, $N_R$ needs to be large, in the order of $N_S$, to fully utilize this advantage. Each segment has the varied number of sampled data points because of the nonlinearity in frequency sweeps as shown in Fig. 4 and their SNR could be affected by the nonuniform ratio $N_R/N_S$ at each segment.

From the discussions above, the SNR in Swept Source Lidar systems $SNR_{SSL}$ in decibels can be written in the form:

$$SNR_{SSL}[dB] = 10\log\left(\frac{\eta_{SSL} P_r(L)T}{E_v N}\right) = 10\log\left(\frac{\eta_{SSL}\eta_t A_r \rho_T P_0 T}{\pi L^2 E_v N}\right), \quad (17)$$

where $\eta_{SSL} = \eta_d \times \eta_s$ is the product of the detection efficiency $\eta_d$ and the transmittance of beams in the system after the receiving aperture $\eta_s$. The estimated SNR by the single FFT signal for $L$ = 4.5 m was 21 dB. Assuming $\eta_{SSL} \times \eta_t = (0.3 \times 0.15) \times (0.89)^2$, the receiving aperture area $A_r = \pi \times (3 \text{ mm}/2)^2$, the target reflectivity $\rho_T = 0.1$, the optical power $P_0 = 1$ mW, the tuning period $T = 800$ μs, and $N = 12$, Eq. (17) predicts a similar SNR of 21 dB. The total loss is a combination of optical losses in the recoupling into the interferometer and fiber-to-fiber connections, decrease of the diffraction efficiency for reflected beams, relative intensity noise of laser light, thermal noise of detectors, and other circuit noises.

The sensitivity can be defined as the inverse of the smallest sample reflectivity where the SNR equals 1 in SS-OCT systems [35]. The sensitivity in Swept Source Lidar systems in decibels can be defined by the smallest reflectance $P_r/P_0$ as written in the form:

$$Sensitivity_{SSL}[dB] = 10\log\left(\frac{\eta_{SSL} P_0 T}{E_v N}\right). \quad (18)$$

High sensitivity is one of the advantages of SS-OCT as the experimental value of 120 dB and theoretical prediction of 126 dB have been reported [22]. Predicted sensitivity of our system by Eq. (18) with given parameters is calculated as 101 dB where the low efficiency $\eta_{SSL}$ and the division of the interference signal into $N$ segments account for the difference of sensitivity from the literature values for SS-OCT systems. To detect a target with $\rho_T = 0.1$ at $L = 200$ m using the receiving aperture radius of 10 mm, a sensitivity of 96 dB is required and the current system achieves this value. Improvement in the total system loss and the use of a longer tuning period $T$ and a higher optical power $P_0$ can compensate the decreased sensitivity with a larger $N$, as will be discussed in Sec. 4.4.

*4.3 Beam steering*

Various wavelength dispersive elements based on mature technologies can be employed in Swept Source Lidar systems. Diffraction gratings, which we have used in the demonstration, have a number of advantages. First, the use of gratings allows us to work with large beam apertures that increase the SNR. Second, the absence of diffraction in the direction parallel to the grating grooves allows flexible optical design [26,32,36]. A grating can be placed after a 1-axis beam scanner, as we have demonstrated, where a large beam steering range can be implemented with a larger grating. A grating may also be placed before a beam scanner considering optical design requirements.

The beam steering range, which is also known as the field of view (FOV), is one of the essential parameters in lidar applications. The whole angular beam steering range $\Theta^{ss}$ is determined by the whole tuning bandwidth $F$ in Swept Source Lidar systems. Assuming a typical tuning bandwidth of 88 nm centered at 1050 nm for our tunable VCSEL [17], the beam steering ranges are calculated and shown in Fig. 9. Though a reflection grating was used in the experiments, a transmission grating is assumed in the calculations because it can separate an incident beam from diffracted beams more easily (Fig. 9(a)).

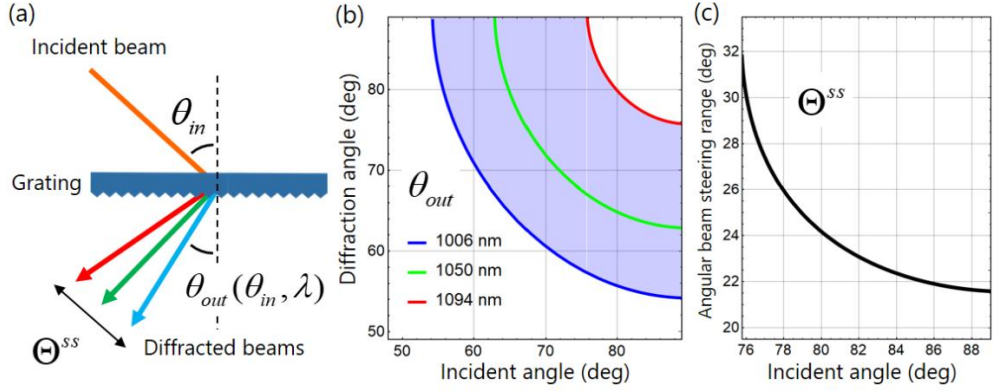

Fig. 9. (a) Schematic of the diffraction of an incident beam with a transmission grating. (b) Calculation of diffraction angles at various wavelengths as a function of the incident angle. The blue, green, and red lines show diffraction angles for the wavelengths of 1006, 1050, and 1094 nm, respectively. (c) Calculation of the whole beam steering range with various incident angles.

The first-order diffraction angles $\theta_{out}(\theta_{in}, \lambda)$ as a function of the incident angle $\theta_{in}$ at various wavelengths $\lambda$ are calculated for a grating with a groove density of 1800 grooves/mm (Fig. 9(b)). At the incident angle $\theta_{in}$ of 76 deg and greater, the entire wavelength sweep range can be utilized for beam steering. The whole angular beam steering range $\Theta^{ss}(\theta_{in}) = |\theta_{out}(\theta_{in}, \lambda_{max}) - \theta_{out}(\theta_{in}, \lambda_{min})|$ for this region is plotted (Fig. 9(c)), where $\lambda_{max}$ and $\lambda_{min}$ are 1094 and 1006 nm, respectively. A vertical FOV of 30 deg, which is typically required in automotive applications, can be achieved with the incident angle $\theta_{in}$ of about 76 deg.

To achieve a larger beam steering range, a beam expander can be placed after the grating. There is a tradeoff to be made between the beam steering range and the SNR. The magnification of the beam steering range in one of two axes by a factor of $M_B$ reduces the SNR by the same factor of $M_B$ as a result of the reduced aperture area $A_r$. To compensate the reduced SNR, a larger original beam aperture may be implemented taking advantage of gratings. Note that the magnification of the beam steering range increases the angular pixel size and can degrade the angular resolution. To keep the angular pixel size, the number of segments $N$ can be increased accordingly.

*4.4 Swept source scan in slow axis*

In order to increase the maximum detection range $R_{max}^{ss}$, a slower frequency ramp rate $F/T$ or a higher sampling rate $B$ is needed. Since $B$ is limited by DAQ systems and a large tuning bandwidth $F$ is required for a large angular beam steering range $\Theta^{ss}$, increasing the tuning period $T$ is a reasonable solution to increase $R_{max}^{ss}$. However, the frame rate decreases as $T$ increases when the swept source scan is used in the fast axis of 2D raster scanning. To overcome this dilemma, the swept source scan can be assigned to the slow axis, so that the frame rate equals to the frequency sweep rate. A higher sensitivity by a longer $T$ allows the system to detect targets over longer ranges and have higher angular resolutions by a large $N$. Thus, assigning the swept source scan to the slow axis provides a solution especially attractive for automotive applications [1,6], where long detection ranges of a few hundred meters, moderate frame rates of 10-20 Hz, and high angular resolutions with large FOVs are demanded.

A schematic of 2D beam steering with the swept source scan used in the slow axis is illustrated in Fig. 10(a). A beam is steered with a diffraction grating in the vertical axis by the swept source scan. During a single swept source scan, the beam repeats fast horizontal scans with a mechanical or solid-state 1-axis beam scanner. In this way, the beam is 2D raster-scanned during a single monotonic frequency sweep. Figure 10(b) depicts the data processing for the

2D raster scanning. First, the obtained interference signal $I_s(t)$ is coarsely divided into $N_V^{ss}$ sections. Then, the signal at $i$th section is finely divided into $N_H^{ss}$ segments. The $j$th segment of the $i$th section corresponds to the signal obtained at the beam steering angle $\theta_{i,j}^{ss}$ (yellow arrow, Fig. 10(a)). The range to a target $R_{i,j}^{ss}$ at $\theta_{i,j}^{ss}$ is obtained from the peak position in the FFT signal (Fig. 10(b)). From the interference signal divided into $N = N_H^{ss} N_V^{ss}$ segments, the ranges to targets over 2D raster-scanning points of $N_H^{ss} \times N_V^{ss}$ are obtained as a 3D point cloud. Because the beam is steered by the beam scanner from each segment to the next, delayed light from different angles cannot return to the system through the grating, unlike the case with the swept source scan assigned in the fast axis.

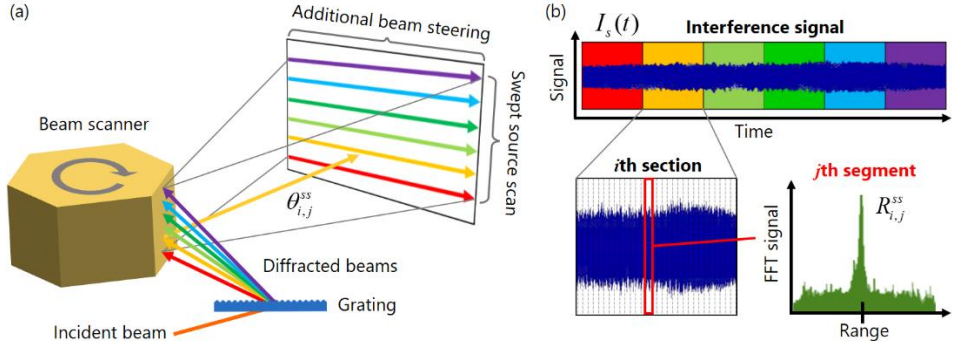

Fig. 10. (a) Schematic of 2D beam steering in Swept Source Lidar system with the swept source scan in the slow and vertical axis, combined with a 1-axis beam scanner. (b) Schematic diagram of data processing. The FFT signal for the $j$th segment of the $i$th section of the interference signal provides the range $R_{i,j}^{ss}$ at the beam steering angle $\theta_{i,j}^{ss}$.

In the following paragraphs, the feasibility for automotive lidar applications is discussed. The swept source scan can realize a typical FOV of 30 deg for the vertical axis. A typical FOV of 100 deg for the horizontal axis can be achieved by a polygon scanner because the maximum optical beam steering angle range of a 6-facet polygon scanner is $(360/6) \times 2 = 120$ deg. When $N = N_H^{ss} N_V^{ss} = 600 \times 45$, angular pixel sizes in the horizontal and vertical directions are 0.17 and 0.67 deg/pixel, respectively. The 2D pixel size at $L = 100$ m is 29 cm (H) × 116 cm (V), which is the moderate pixel size to detect pedestrians, cars, and the surrounding environment. When the frequency sweep rate is set to a typical frame rate of 10 Hz, $T$ of 80 ms can be assumed. This gives the required rotation speed for a 4-facet polygon scanner of $(45/80[\text{ms}]) \times (60/4) = 8{,}400$ RPM, which can be realized [37]. If a 1-axis MEMS mirror is used, a scanning frequency of $45/80[\text{ms}] = 560$ Hz is required. Because MEMS mirrors typically have limited optical beam steering angle range (e.g., 52 deg at 500 Hz [38]), multiple MEMS mirrors can be combined to cover large FOVs.

Assuming the sampling rate $B = 1$GS/s and a typical whole tuning bandwidth $F = 24$ THz of our tunable VCSEL achieving 30-deg beam steering, the maximum detection range $R_{\max}^{ss}$ calculated by the tuning period $T = 80$ ms is 240 m, which is a typically expected maximum detection range. The number of sampling data points $N_S$ at each segment is $BT/N = 3{,}000$ on average and $N_R$ can be set somewhat larger than $N_S$ to be $4{,}096$ ($= 2^{12}$). This gives the whole detection range $Z_{FFT}^{ss}$ of 340 m and is longer than $R_{\max}^{ss}$. The theoretical resolution $\Delta R_i^{ss}$ calculated by the tuning bandwidth for each segment $F/N = 0.9$ GHz is 17 cm, which is comparable to the range pixel size $\Delta Z_{FFT}^{ss}$ of 17 cm, is sufficient to detect the environment. If $\eta_{SSL}$ is improved to $(0.3 \times 0.6) \times (0.89)^2$ and $P_0 = 5$ mW is used with our tunable VCSEL, the sensitivity predicted from Eq. (18) is 101 dB, surpassing 96 dB, which is required to detect a target with $\rho_T = 0.1$ at $L = 200$ m. These calculations verify that a large FOV (100 deg (H) × 30 deg (V)) with fine pixel sizes (0.17 deg (H) × 0.67 deg (V)) and a long detection range of 200-240 m with a high range resolution of 17 cm at a frame rate of 10 Hz can be achieved with Swept Source Lidar systems using the swept source scan in the slow axis.

## 5. Conclusion

In conclusion, we have proposed "Swept Source Lidar", which realizes FMCW ranging and nonmechanical beam steering simultaneously and continuously using wavelength dispersive elements. Our scheme allows us to eliminate the need to steer beams along at least one of the two axes of 2D beam steering, simplifying the FMCW lidar system as a whole. Swept Source Lidar systems can be combined with other beam steering mechanisms. Applying our SS-OCT technology to our FMCW lidar systems, we have demonstrated our scheme using a tunable VCSEL as a wideband swept source. We believe that our tunable MEMS-VCSEL is suitable for Swept Source Lidar systems because of its wide tuning bandwidth and the long coherence length. Employing a galvo scanner as a mechanical beam steering mechanism along one of the two axes, 3D Swept Source Lidar data has been successfully obtained over 2D beam steering at a high frequency sweep rate of 10 kHz. Furthermore, our scheme for a long range of 5 m has been demonstrated in 3D with a slow frequency sweep rate of 300 Hz.

We have also discussed the feasibility, scalability, and limitations of Swept Source Lidar systems, regarding the SNR, the detection range, and beam steering mechanisms. By using the swept source scan in the slow axis of 2D beam steering, the typical requirements for automotive lidar applications can be achieved with Swept Source Lidar systems. To realize a full solid-state FMCW lidar system, swept source scans can be used along both axes of 2D beam steering, or various 1-axis solid-state beam steering techniques can be combined in a Swept Source Lidar system. Thus, Swept Source Lidar systems that we have proposed and demonstrated in this work can lead to the realization of simple solid-state FMCW lidar systems.

## Acknowledgments

The authors would like to thank Prof. Nobuhiko Nishiyama (Tokyo Institute of Technology), Prof. Hiroshi Toshiyoshi (the University of Tokyo), Mohammed Saad Khan, Kiyoteru Nabeno, Yi Xiao, Chang-Dae Keum, and Keiji Isamoto (Santec Corp.) for their supports on tunable VCSELs, and Takuya Suzuki (Santec Corp.) for software development.

## Disclosures

The authors declare no conflicts of interest.